\documentclass{aa}
\usepackage{graphicx}

\begin{document}

\newcommand{\CM}{~cm$^{-1}$~}
\newcommand{\MU}{[$\mu$]~}
\newcommand{\CC}{$^{12}$C$^{16}$O~}
\newcommand{\CCC}{$^{13}$C$^{16}$O~}
\newcommand{\CDC}{$^{12}$C/$^{13}$C~}
\newcommand{\EEE}{~E$^{``}$~}
\newcommand{\Tef}{T$_{\rm eff}$~}
\newcommand{\HHO}{H$_2$O~}
\newcommand{\DG}{$\Delta_t$~}
\newcommand{\mum}{$\mu$m~}

\title{Carbon Monoxide bands in M dwarfs}

\author{Yakiv V. Pavlenko\inst{1} \and Hugh R.A. Jones\inst{2}}

\offprints{Ya.V. Pavlenko}
\mail{yp@mao.kiev.ua}

\institute{Main Astronomical Observatory of Academy of Sciences of
Ukraine, Golosiiv woods, Kyiv-127, 03680 Ukraine, e-mail:
yp@mao.kiev.ua
\and Astrophysics Research Institute, Liverpool John Moores University,
Egerton Wharf, Birkenhead CH41 1LD, UK
e-mail: hraj@astro.livjm.ac.uk}

\date{Received ; accepted }

\authorrunning{Yakiv V. Pavlenko and Hugh R.A. Jones}
\titlerunning{CO bands in M dwarf spectra}

\abstract{We compare the observational and theoretical spectra of
the $\Delta v$ = 2 CO bands in a range of M dwarfs. We investigate
the dependence of theoretical spectra on effective temperatures as
well as carbon abundance. In general we find that the synthetic
CO bands fit the observed data extremely well and are excellent
diagnostics. In particular the synthetic spectra reasonably match
observations and the best fit temperatures are similar to
those found by empirical methods. We also examine the \CDC
isotopic ratio.
We find that fundamental $^{13}$CO bands around 2.345 and 2.375 $\mu$m
are good discriminators for the \CDC ratio in M dwarfs.  
The 2.375 $\mu$m is more useful because it doesn't suffer such serious
contamination by water vapour transitions. Our current dataset 
does not quite have the wavelength coverage to perform a reliable
determination of the  \CDC ratio in M dwarfs. For this we recommend
observing the region 2.31--2.40 $\mu$m at a resolution of better than 
1000. Alternatively
the observational problems of contamination by water vapour at 
2.345 $\mu$m maybe solved by observing at resolutions of around 50000. 
We also
investigated the possibility of using the $\Delta v$ = 1 CO bands
around 4.5 $\mu$m. We find that the contamination due
to water vapour is even more of a problem at these wavelengths.
\keywords{energy
distributions -- Stars: effective temperatures}} 

\maketitle

\section{Introduction}
More than 70\% of stars in the vicinity of the Sun are M dwarfs.
These numerous low-mass stars (0.08 M$_{\odot}$ $ \leq $ M $ <
$ 0.6 M $_{\odot} $), together with substellar objects - brown
dwarfs (M $ \leq $ 0.08 M$_{\odot} $) can contain an appreciable amount
of the baryonic matter in the Galaxy. Research of M dwarf
spectra are of interest for many branches of modern astrophysics.
Verification of the theory of stellar evolution and structure of
stars, the detection among M dwarfs of a subset of young brown dwarfs,
the physical state of plasma of their atmospheres at low
temperatures, as well as the chemical and physical processes of dust
formation are only a few of them.

The dominant opacity sources in the optical and IR spectra of M dwarfs
is absorption by band systems of diatomic molecules, such as TiO and VO,
as well as rotational-vibrational bands of  \HHO. M dwarf infrared spectra
additionally contain absorption bands of CO.
Apart from being readily identifiable in M dwarf spectra, CO bands
have the advantage of being extremely well modelled. This is
in contrast to the relatively poor modelling of other molecules
and atomic lines in M dwarf atmospheres.
For example, the structure of
rotational-vibrational levels of two-nuclear molecules CO
is much simpler, than \HHO. That allows computations
of CO line lists to be carried out with high accuracy (e.g., Goorvitch 1994).
One of the most promising observational
regions is located in the K band from 2.2 to 2.4 $\mu$m. Second
overtone bands \CC and \CCC are located here. As well as parameters
such as effective temperature and gravity, they can be used for
determination of carbon and/or oxygen abundances and the \CDC
ratio in atmospheres of late-type stars.
The determination of the \CDC ratio in M dwarf atmospheres is especially
interesting.
Following the conventional theory of
stellar evolution (see Aller \& McLaughlin 1965), M dwarfs save
their initial \CDC from their time of formation. Since the galactic
\CDC ratio is expected to change by around a factor of four over 
the lifetime of our Galaxy,  
the determination of the \CDC ratio for M dwarfs potentially
gives a strong constraint on their age. 
However, determining the 12C/13C ratio is
only a strong constraint on age if that ratio is a single-valued
function of time and Galactic location. If that underlying
assumption is not valid, any dispersion of \CDC in M-dwarf
atmospheres might give for us
some clues about mixing processes inside our Galaxy.
Until now, measurements  
of the \CDC ratio 
for planetary nebulae, red giants and asymptotic giant
branch stars do not fit particularly well with theory.
Nonetheless, the discrepancies found for such evolved stars may 
well arise from the lack of realistic stellar models. Atmospheric 
models need to include
additional physical processes in their prescription for mixing between  
nucleosynthetic cores and observable atmospheres (Palla et al. 2000). 
M dwarfs, on the other hand, are not expected to modify their 
\CDC and are fully convective.
Thus the \CDC ratios for a diverse sample of M dwarfs is expected
to be a relatively straight-forward function of time and galactic 
location. 

Until now studies of CO bands in the K-band were mainly used for the
quantitative analysis of the evolution of \CDC ~in atmospheres of red
giants (cf. Hinkle 1978; Sneden \& Pilachowski 1984;
McGregor 1987; Lazaro et al. 1991; Suntzeff \& Smith 1992;
Scott et al. 1994;
Pilachovski et al. 1997). However, they are also observed in the spectra
of many objects - from symbiotic stars (Schild et al. 1992), Be
stars (Kraus et al. 2000) and supernova remnants
(Gerardy et al. 2000) up to planets (Forbes et al. 1970).
Baldwin et al. (1973) and many others since have shown the prevalence 
of the CO bands in the K band spectra of M dwarfs. Viti et al. (2002)
have shown that the CO bands in the K band are relatively well-matched by
synthetic spectra and thus can be used to determine M dwarf properties.   
In our study we make comparisons of synthetic and observed spectra
across the 2.3--2.4 $\mu$m region to quantify the usefulness of the CO
bands as a diagnostic of effective temperatures and metallicities
for M dwarfs.
\section {Observations}

\begin {table*}
\caption {The properties of observed stars are taken 
from Jones et. al. (1996). Spectral types are taken from Kirkpatrick, Henry
\& McCarthy (1991)}
\begin {tabular} {ccccccccc}
\hline
\hline
\noalign{\smallskip}

Object & $d$ (pc) & KIN & COL & Sp Type & $M\rm _K$(CIT) &
$V\rm _C$--$K\rm_{CIT}$ & $I\rm _{CCD}$--$L'\rm _{MKO}$ \\
\noalign{\smallskip}
\noalign{\smallskip}

\hline

\noalign{\smallskip}
Gl~411 & 2.5 & OD & OD/H & dM2     & 6.34 & 4.11 & 2.24\\

Gl~699  & 2.39 & OD/H & OD/H & dM4  & 8.20 & 5.04 & 2.59 \\

Gl~406   & 1.83 & OD & - & dM6     & 9.19 & 7.37 & 3.70 \\

VB~10  & 5.79 & OD & - & dM8       & 9.99& 8.70 & 4.65\\

LHS~2924  & 10.72 & OD & - & dM9   &10.52 & 8.91 & 5.19 \\

\hline

\end{tabular}
\end{table*}

The properties of observed stars for which spectra are fitted in our paper
are listed in Table 1.
Observations were made with the Cooled Grating Spectrometer 4
on the UK Infrared Telescope (UKIRT) on Mauna
Kea, Hawaii.  The instrument then had a 58 $\times$ 62 InSb array which was
moved in the focal plane in order to three times over-sample the
spectrum.  Sky subtraction was performed by nodding the telescope
approximately 30 arcsec up and down the slit, ensuring that during
alternate `object' and `sky' observations the star remained on the
detector.  The observations presented in this paper were made during
three nights, 1992 May 8, 10 and 12 in reasonable optical seeing
(0.75--1.5 arcsec) and of
atmospheric humidity (10--50 per cent).  The airmass difference between
object and standard never exceeded 0.05 and so we are confident that the
spectra have good cancellation of atmospheric features.

The 150 lines mm$^{-1}$ grating was used in third order with the 150 mm
focal length camera at a central grating wavelength of 2.34 $\mu$m.  This
grating position was chosen to be closed to the centre of the
CO bands, at a maximum in the grating efficiency have
high atmospheric transmission ($>$95 per cent) and gives a relatively high
resolution, $\lambda/\Delta\lambda$, where $\Delta\lambda$ is the
detector resolution at wavelength $\lambda$, of 1085 (equivalent to 276
km/s).

\subsection{Standards} \label{s2-stds}

Stars in the spectral type ranges B5--A5 and F6--G0 were used to remove
the effects of atmospheric absorption.  These standards were expected to
be featureless at the spectral resolution used and to be well described
by a Rayleigh-Jeans tail.  However, the F6--G0 standards were not very
useful as they show weak metal lines.  Although the lines
tended to be weaker and different from the M dwarf features we avoided using
them where possible.
In the reduction of spectra, we have followed the same procedures as in
Jones et al. (1996).
All observations were wavelength calibrated using lines from
observations of an argon lamp in the CGS4 calibration unit.  This
procedure is typically accurate to 0.1 $\Delta$$\lambda$.

\section{Procedure}

We used local thermal equilibrium (LTE) model atmospheres of M
dwarfs with effective temperatures \Tef = 2400--3800 K from the
NextGen grid of Hauschildt et al. (1999). Unless otherwise
mentioned all models are for log g = 5.0 and solar metallicity.
Computations of LTE synthetic spectra were carried out by the
program WITA6 (Pavlenko 2000) assuming LTE, hydrostatic
equilibrium for an one-dimensional model atmosphere and without 
sources and sinks of energy. The equations of
ionisation-dissociation equilibrium were solved for media
consisting of atoms, ions and molecules. We took into account
$\sim$ 100 components (Pavlenko 2000). The constants for equations
of chemical balance were taken from Tsuji (1973). We adopt the
dissociation potential D$_{\rm 0}$ =9.5 eV for \HHO and D$_{\rm 0}
$ =11.105 eV for \CC, \CCC. It is worth noting that the chemical
balance in M dwarf atmospheres is governed by the CO molecule (see
section 4.1).

Line lists of $^1$H$_2^{16}$O were computed using the AMES
database (Partrige \& Schwenke 1998). The partition functions  of
\HHO ~were also computed on these data (Pavlenko 2002). We used \CC
and \CCC ~line lists of Goorvitch (1994). The CO partition
functions were taken from Gurvitz et al. (1989). The atomic line
list was taken from VALD (Kupka et al. 1999).

The profiles of molecular and atomic lines are determined using
the Voigt function $H(a,v)$, parameters of their natural broadening
$C_2$ and van der Waals broadening $C_4$ from databases (Kupka et
al. 1999) or in their absence computed following Unsold (1955).
Owing to the low temperatures in M dwarf atmospheres and
consequently, electron densities, Stark broadening may be
neglected. As a whole the effects of pressure broadening prevail.
Computations for synthetic spectra were carried out with a step
0.00005 $\mu$m for microturbulent velocity $v_t$ = 2 km/s.

The instrumental broadening was modelled by triangular profiles set
to the resolution of the observed spectra. To find the best fits to
observed spectra we follow the scheme of Jones et al. (2002).
Namely, for every spectrum we carry out the minimisation of a 3D function
$S=f(x_s, x_f, x_w) = 1/N \times \sum(1-F_{obs}/F_{synt})^2$, where $F_{obs},
F_{synt}$ are observed and computed fluxes,{\em N} is the number of points in
observed spectrum to be fitted,
 $x_s, x_f, x_w$ are
relative shift in wavelength scale, a normalisation factor which
was used to coincide observed and computed spectra and parameter
of instrumental broadening, respectively.

\section {Results}
\subsection {CO and \HHO molecules in M dwarf atmospheres}

The overall chemical balance in the atmospheres of cool stars depends
significantly on the  C/O  = log N(C) - log N(O)(see Tsuji 1973).
In the atmospheres of cool (\Tef $<$ 3600 K) oxygen rich stars
C/O $ < $ 1, almost all atoms of carbon are bound in CO molecules.
CO and \HHO are the most abundant molecules containing
oxygen atoms.

\begin{figure}
\begin{center}
\includegraphics [width=88mm, height=50mm, angle=00]{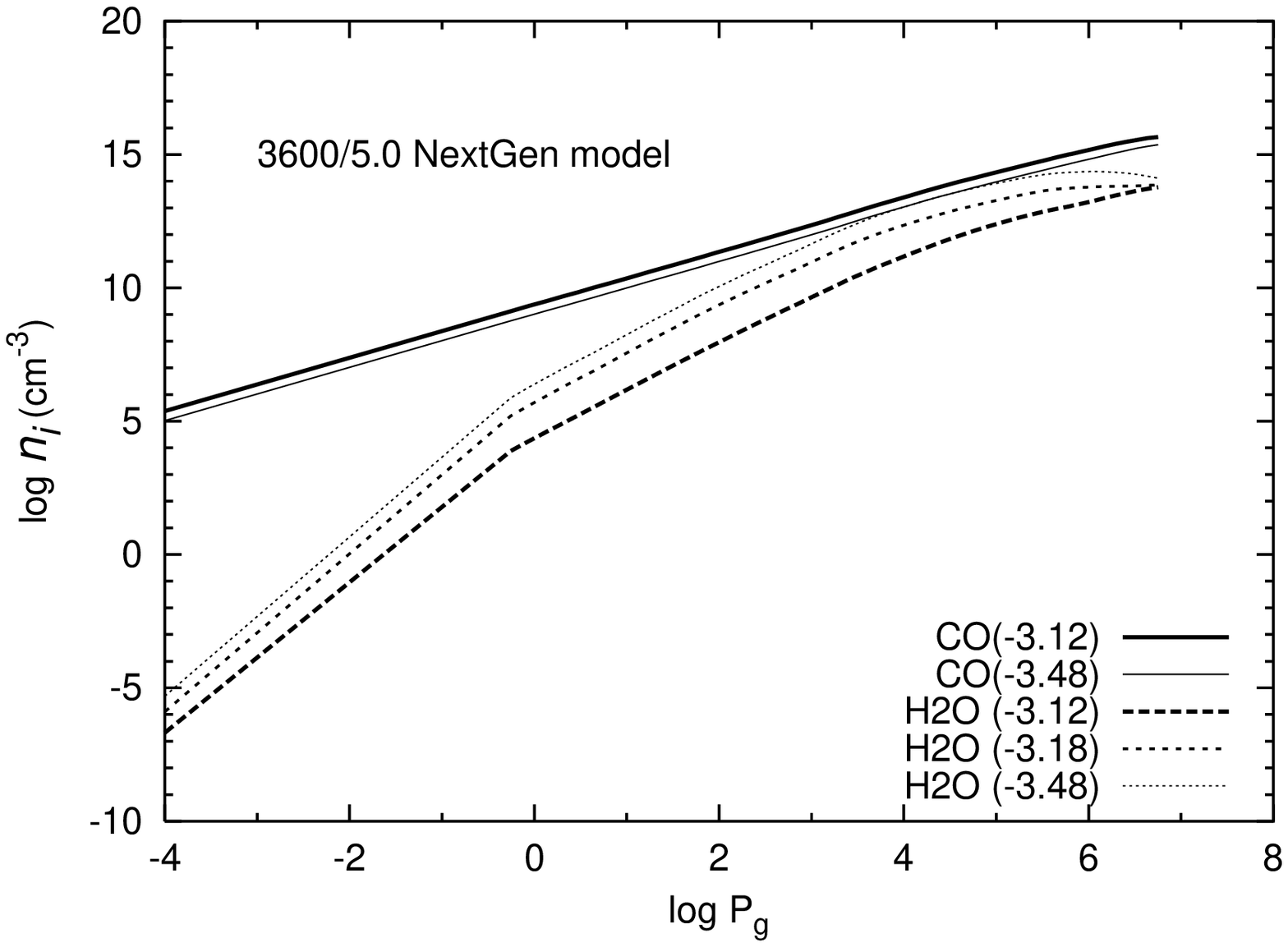}
\includegraphics [width=88mm, height=50mm, angle=00]{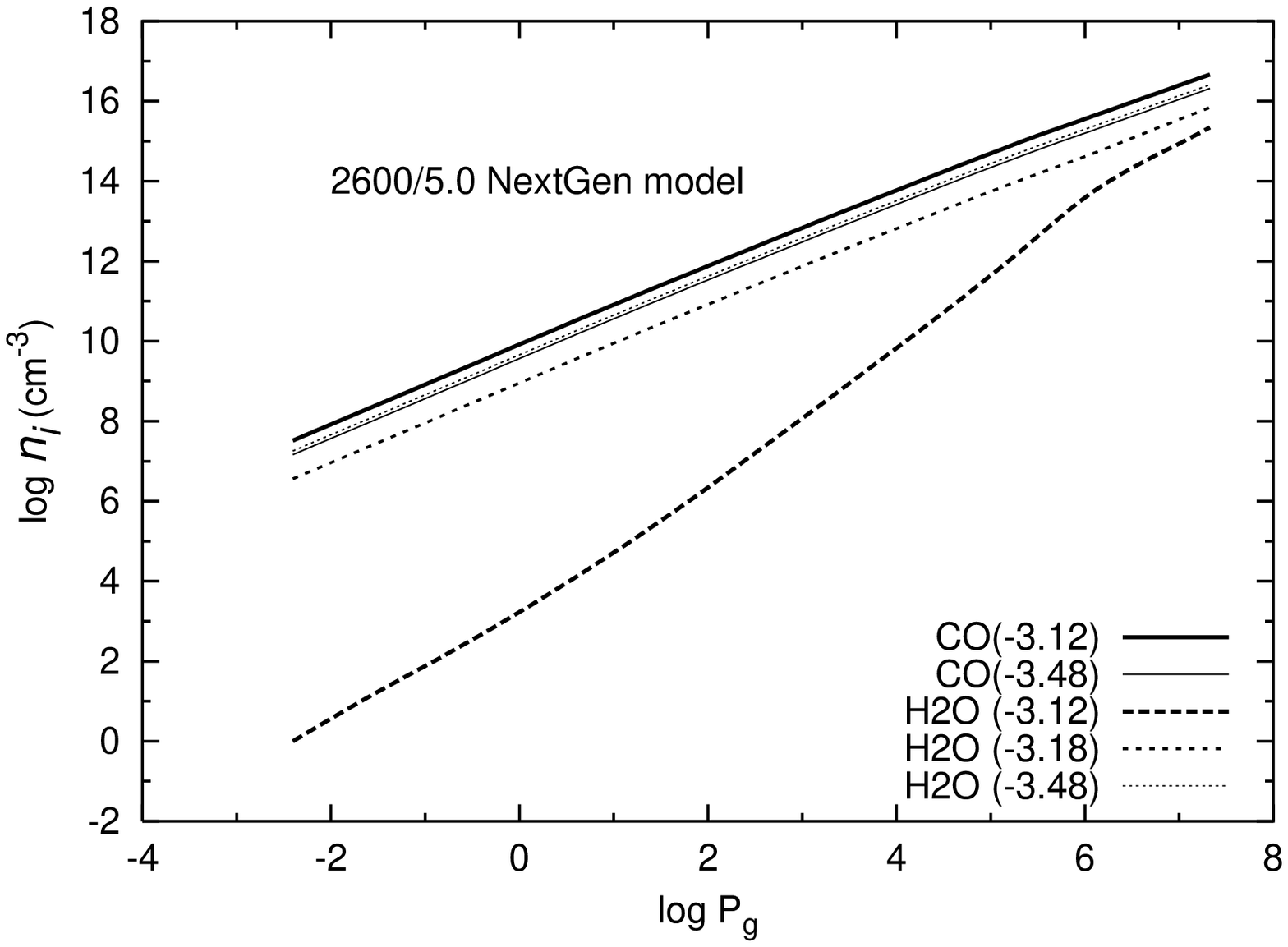}
\end{center}
\caption[]{\label{_MD_} Molecular densities of \HHO and CO in
atmospheres of M dwarfs 3600/5.0 and 2600/5.0 as a function of
gas pressure P$_g$. The case of log N(C) = -3.48 and log N(O) = -3.12
represent the solar abundances case. }
\end{figure}

In the atmospheres of M dwarfs with \Tef $>$ 2600~K, 
molecules of CO dominate by number in comparison with other oxygen
containing molecules (Tsuji 1973). 
Throughout most of the atmosphere 
changes of the ratio C/O  by $\pm$ 0.2 dex have little effect on
the molecular densities of CO (Fig. \ref{_MD_}). In
contrast the dependence of molecular densities on \HHO on log N
(C) is  more complex. At lower
\Tef (e.g. Fig. 1b) the dependence of $n_{\rm H_2O}$ on log N(C) in M dwarfs
atmospheres weakens appreciably though the chemical balance
remains governed by CO. In general, \HHO and CO are
competitors for bound-free oxygen atoms. Therefore their molecular
densities respond in opposite ways to changes of carbon
abundance. For solar abundances of oxygen and carbon
molecular densities of \HHO and CO are comparable. CO
molecular densities increase only a little when log N(C) increases
from -3.48 up to -3.12\footnote{In our paper we use an abundance scale
$\Sigma N_i = 1$.}. At the same time CO reduces molecular
densities of \HHO. However, relative changes of \HHO densities are
comparatively weak for variations of log N(C) in the range from -3.48
to -3.18. Only when the carbon abundance rises to log N(C) $\sim$
-3.18, molecular
densities of \HHO decrease, as an appreciable fraction of oxygen
atoms become bound in CO.

\begin{figure}
\begin{center}
\includegraphics [width=88mm, height=50mm, angle=00]{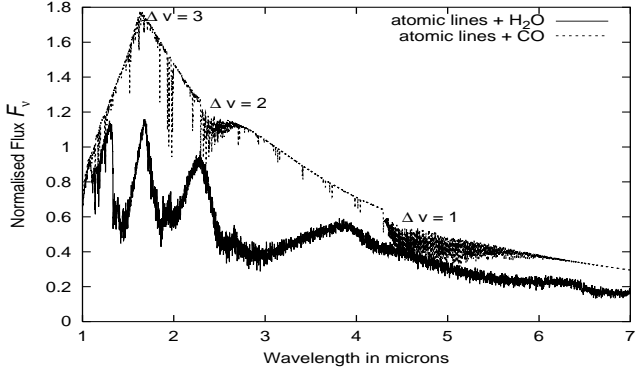}
\end{center}
\caption[]{\label{_H2O_CO_} The relative strength of \CC and \HHO
bands in a 2600 K synthetic spectrum. The positions and
relative strengths of the CO bands can be seen.}
\end{figure}

The relative strengths of \HHO and CO bands at 2600 K
and 3600 K are shown in Fig.\ref{_H2O_CO_}. These spectra were
computed separately by switching on/off the absorption of \HHO or
CO bands.

The relative intensity of the CO bands decrease
from $\Delta v$  = 1 to
$\Delta v$ = 3. Although the second overtone bands ($\Delta v $ = 2)
of CO have intermediate intensities
they are located in a relatively accessible region for observation.

\subsubsection{Bands of the second overtone
of CO from 2.28--2.4 $\mu$m}

\begin{figure}
\begin{center}
\includegraphics [width=88mm, height=50mm, angle=00]{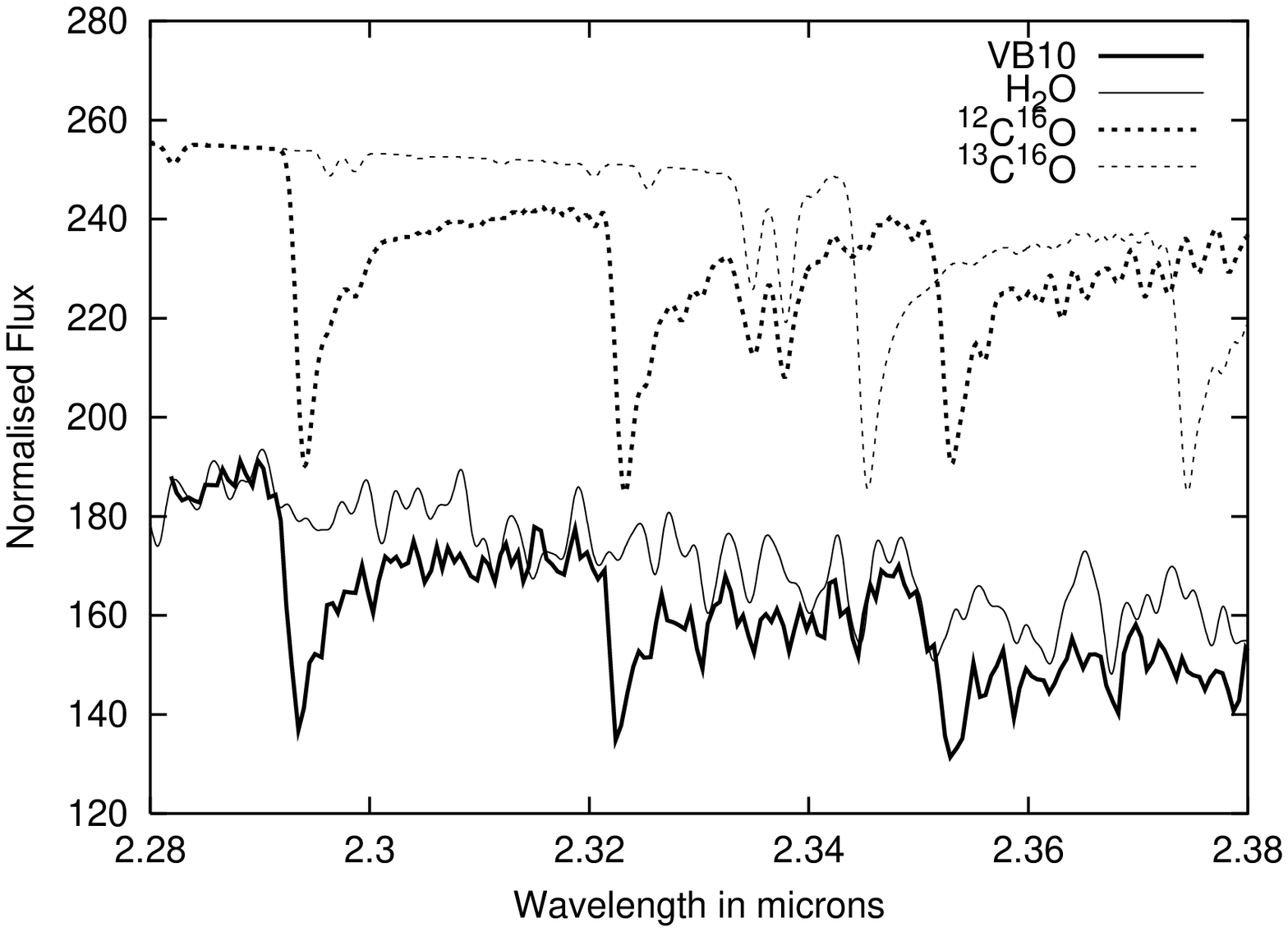}
\includegraphics [width=88mm, height=50mm, angle=00]{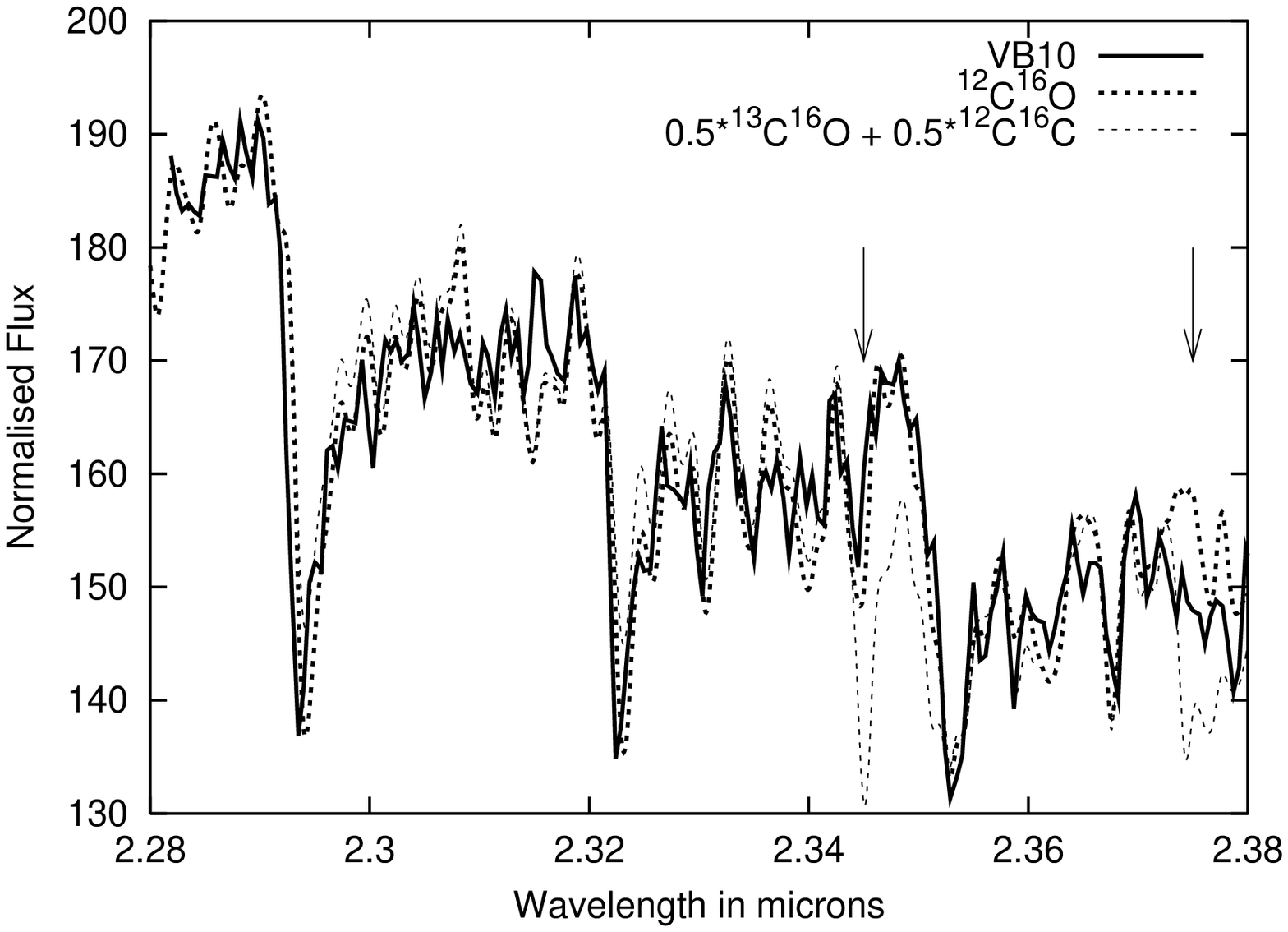}
\end{center}
\caption[]{\label{_CCHHO_} The relative strength of \CC , \HHO and
\CCC bands in the spectrum of VB10 (top), and a combined spectra
computed for the case of \CC + \HHO and \CC + \CCC + \HHO
absorptions. The model atmospheres shown are
2600 K, \CDC = 0.5(bottom).
The positions of \CCC bands heads are shown by arrows.
}
\end{figure}

The relative intensities of \CC, \CCC and \HHO bands from 2.28--2.38
$\mu$m are shown in  Fig. \ref{_CCHHO_}. These computations are
carried out for a model atmosphere with 2600 K and \CDC = 0.5. The
large ratio of \CDC was chosen to show the positions of the main
details which might be observed in M dwarf spectra.
Overall, \HHO absorption dominates the infrared spectral region. However, 
strong features are also formed there by first overtone bands of \CC. 
The positions of the \CCC bandheads
are marked by vertical arrows in Fig. 3. The results of our \CCC
modelling of spectra of our dwarfs are discussed in sections 4.1.3 and
4.1.4.
 
\subsubsection {Fits to the observed spectra}

Our numerical procedure for the determination of best fits
allows us to quantify the possible solutions. For each
observed spectra  we compute synthetic spectra for
different \Tef. We also allow for small
variations of carbon abundances. Based on Fig. 1,
we suggest that small variations of log N(C) from -3.48 to -3.28
cannot
substantially affect the temperature structure of the model
atmospheres.
From the family of curves $S= f$(\Tef, logN (C)) we
choose ``the best solution'' for
each observed spectrum.

\begin{figure}
\begin{center}
\includegraphics [width=88mm, height=50mm, angle=00]{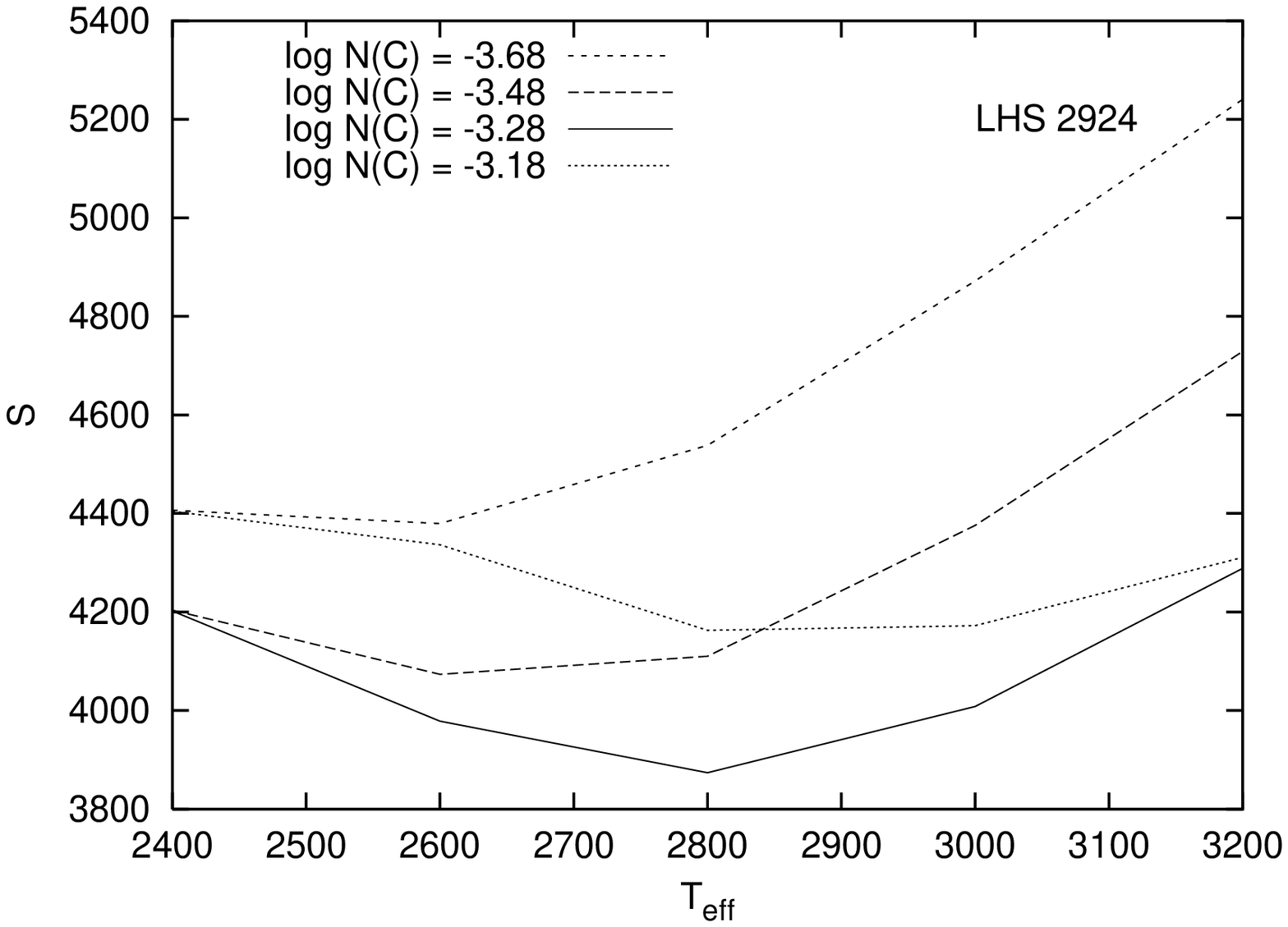}
\includegraphics [width=88mm, height=50mm, angle=00]{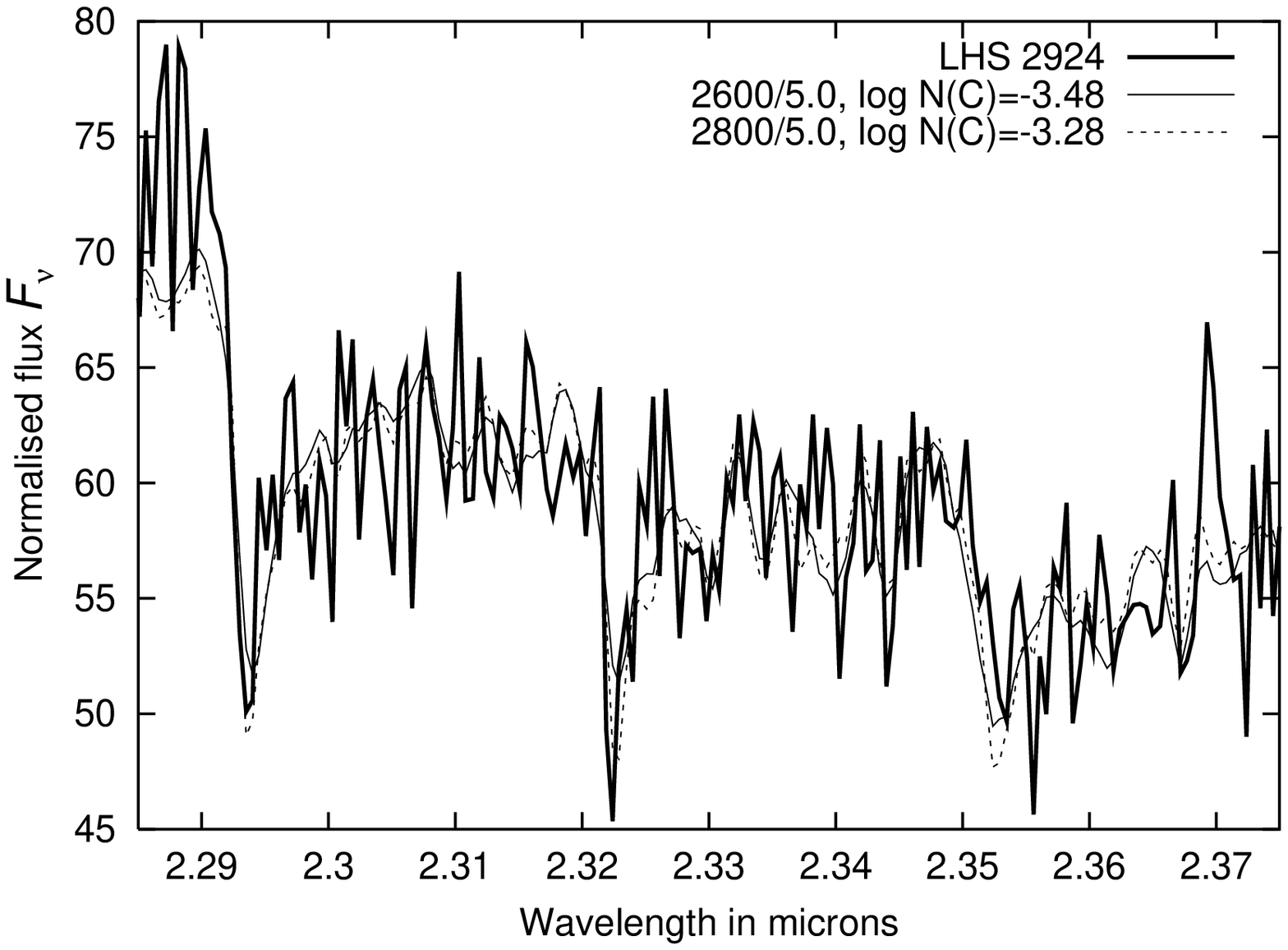}
\end{center}
\caption[]{\label{_LHS2924_}
Top: The dependence of $S$ on \Tef and log
N(C). Bottom: 2600 and 2800 K synthetic spectra compared to the observed
LHS2924 spectrum.}
\end{figure}

{\large\bf LHS2924}. The observed spectrum of this M9V object
is of relatively low signal-to-noise. For example, 
our best fit minimisation value $S_{\rm min}$ for LHS2924
is a factor 10 larger than VB10 (fig. \ref{_LHS2924_}).
Our best fit provides \Tef =
2800 K for log N(C) = -3.28 and \Tef = 2600 K for a solar
abundance of carbon (log N(C)=--3.48). 

{ \large\bf VB10}. VB10 is an archetypal
late-type M dwarf with an effective temperature \Tef $ \sim $ 2600 K
(Jones et al. 1996). Lithium lines are absent from VB10's
spectrum (Schweitzer et al. 1996; Martin 1999) and it has typical old disk
properties (Leggett 1992) and hence VB10 is a star right at the
end of the main sequence.
For the solar abundance of carbon and other
elements $S_{\rm min}$ occurs for 2600 K. However, if we increase
log N(C), $S$ decreases, and we find a best fit value of
2800 K for log N(C) = -3.28$\pm$0.1. In general, 
for higher log N(C), CO
bands intensities increase and best fits move to higher \Tef.
The same effect is found for all dwarfs of our sample.

\begin{figure}
\begin{center}
\includegraphics [width=88mm, height=50mm, angle=00]{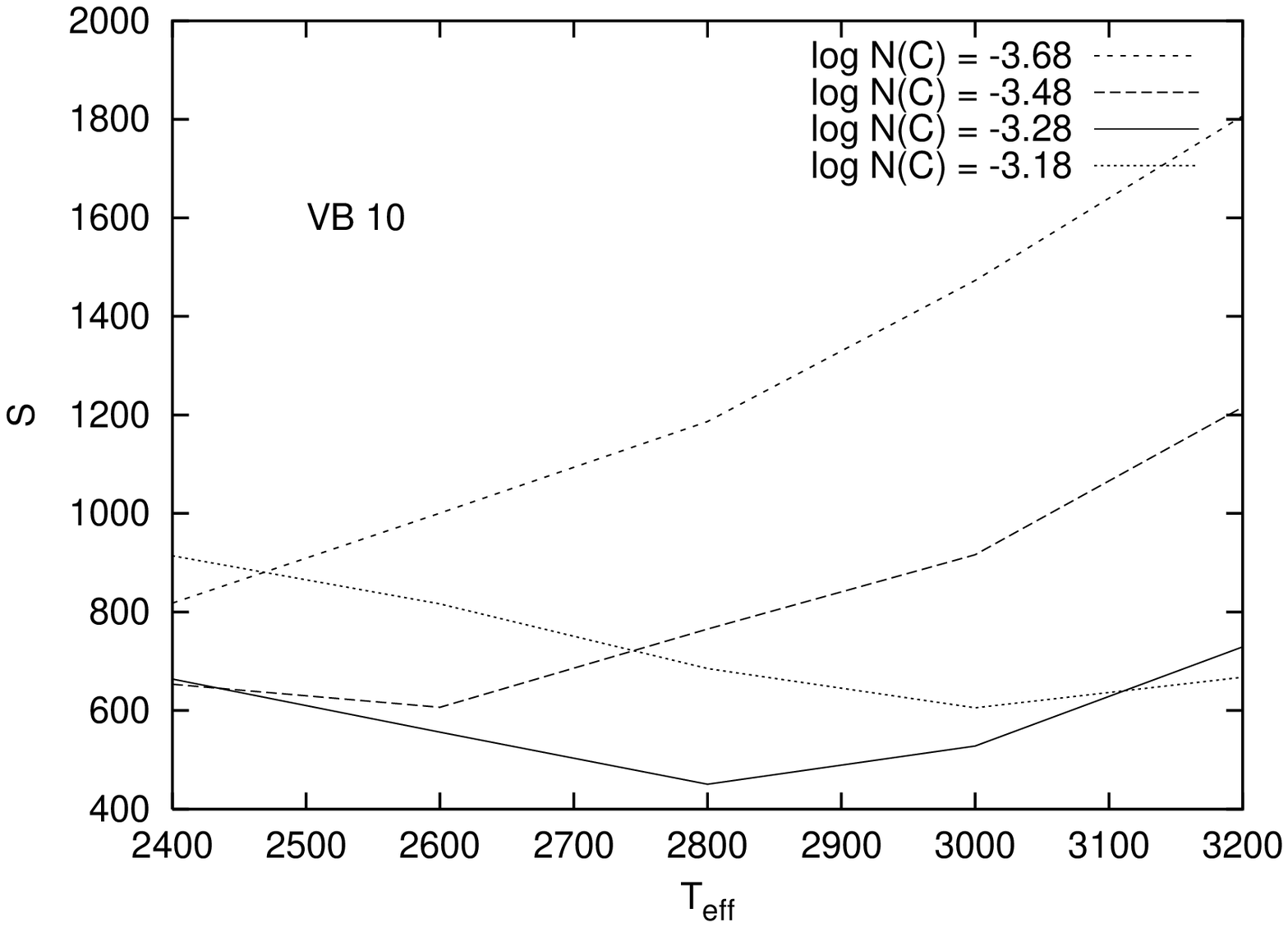}
\includegraphics [width=88mm, height=50mm, angle=00]{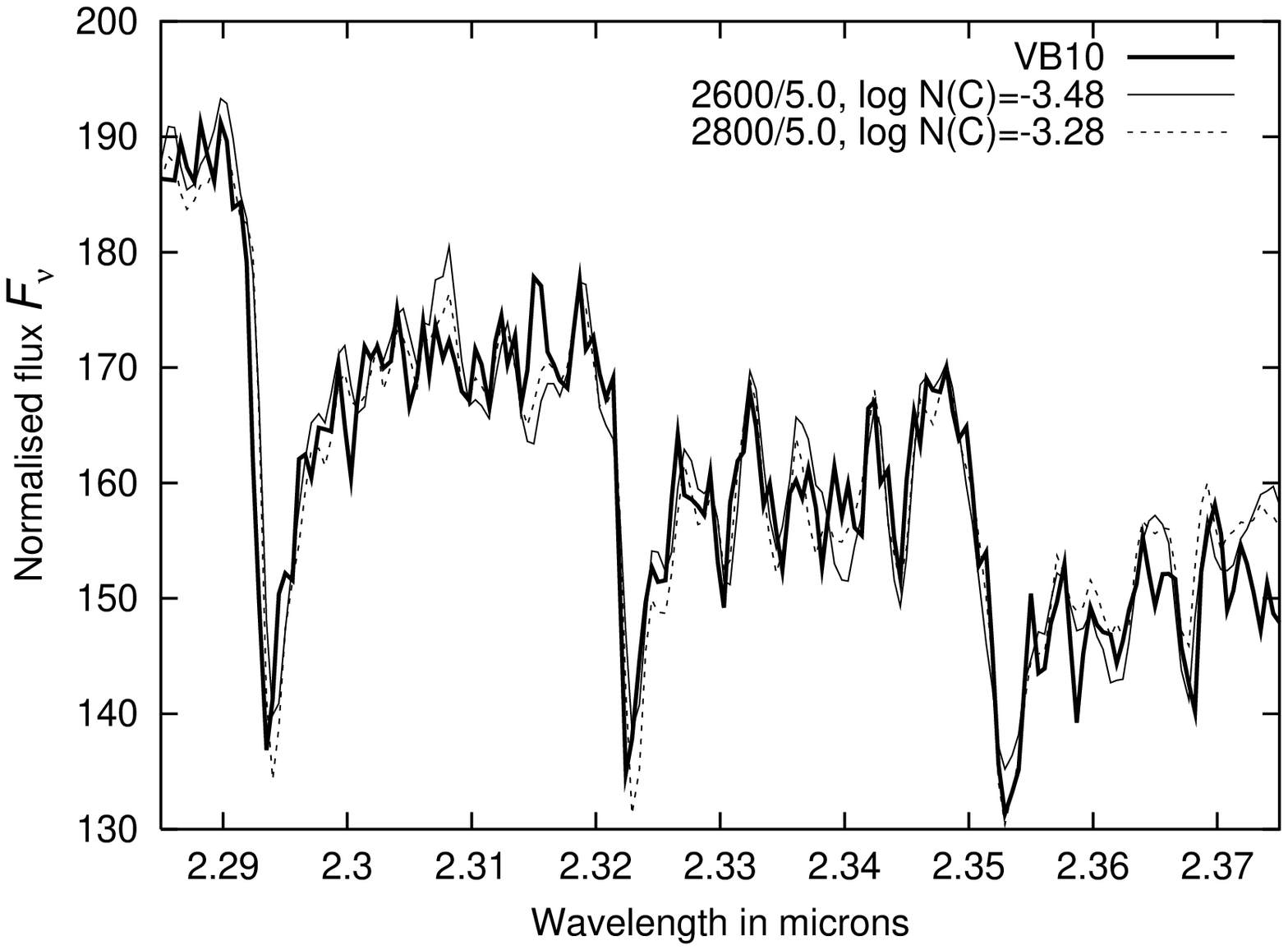}
\end{center}
\caption[]{\label{_VB10_} Top: The dependence of $S$ on \Tef and log
N(C). Bottom: The best fit of synthetic spectra to the observed
VB10 spectrum.}
\end{figure}

\begin{figure}
\begin{center}
\includegraphics [width=88mm, height=50mm, angle=00]{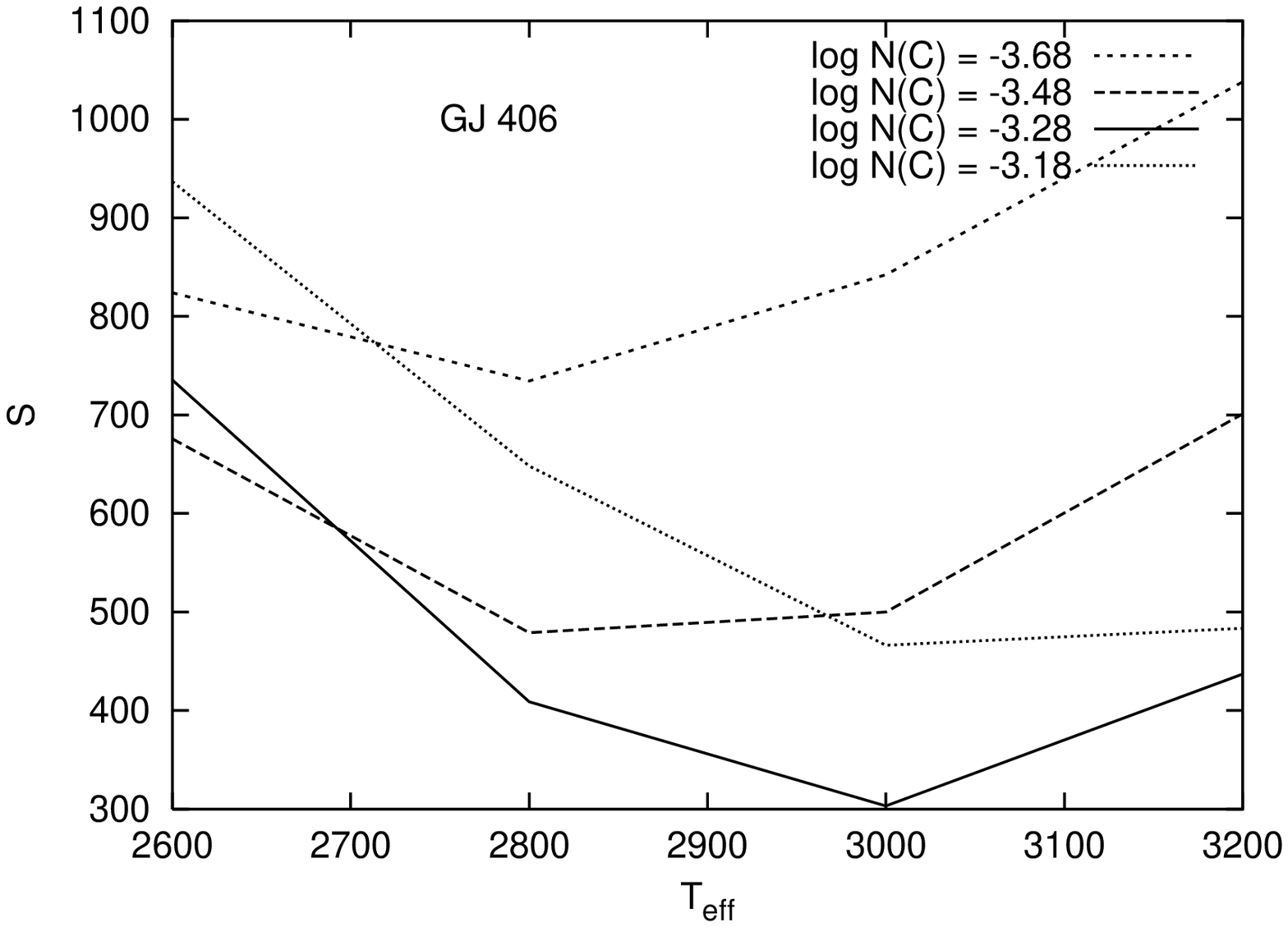}
\includegraphics [width=88mm, height=50mm, angle=00]{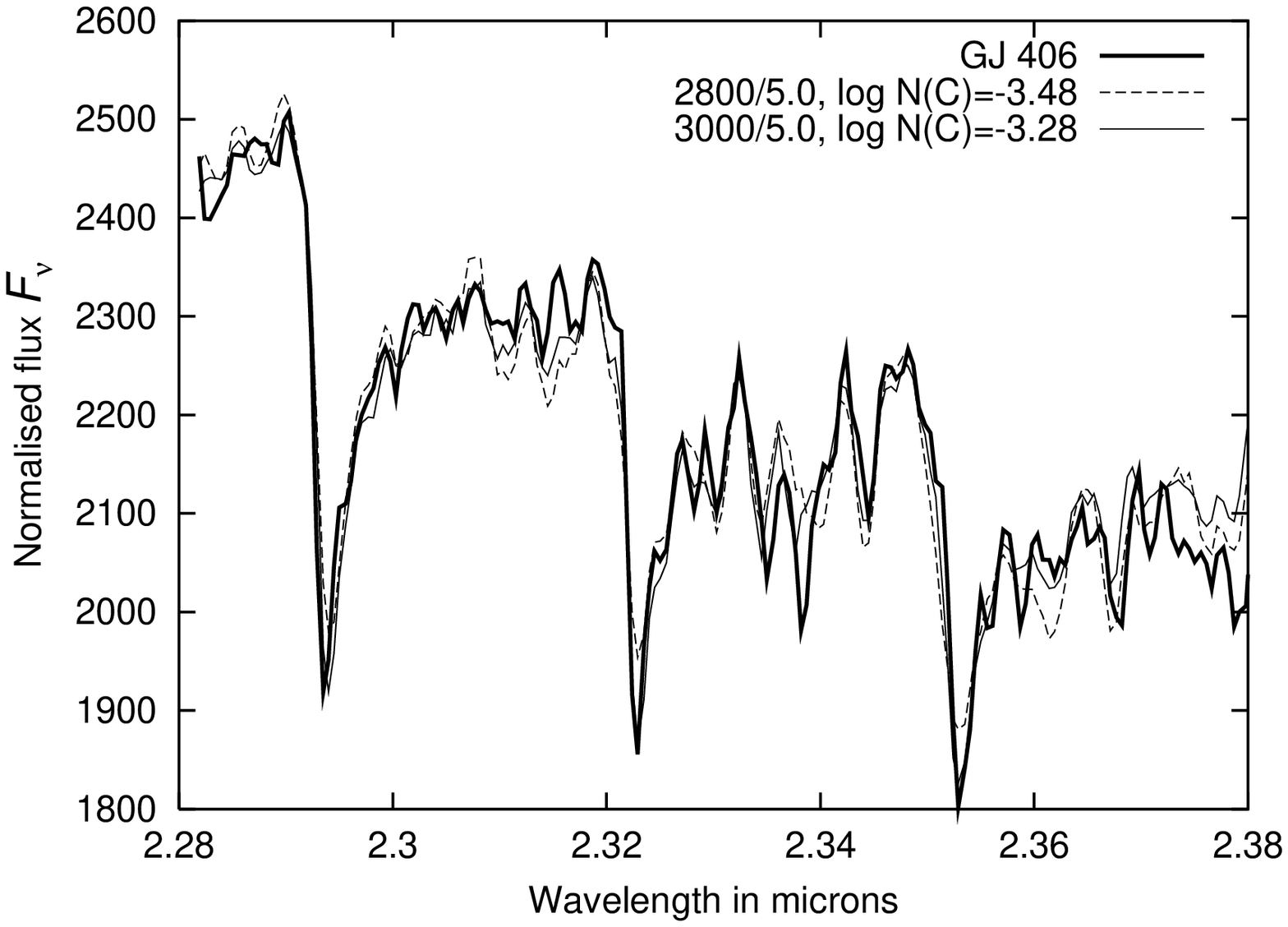}
\end{center}
\caption[]{\label{_GJ406_} The same as in Fig. \ref{_VB10_}, but
for GJ406. Synthetic spectra are shown for \Tef =
2800 and 3000 K.}
\end{figure}

{\large\bf GJ406}.
Best fits are found for 2800 K for the solar abundance case,
and 3000 K for log N(C) = -3.28 (Fig.
\ref{_GJ406_}). Our new estimation of effective temperature
corresponds better with empirical values for the effective
temperature of the spectral class M6V than the Jones et al.
(2002) analysis found using a similar technique but using a
region dominated by water vapour. Our best fits for GJ406 are found
for solar metallicity rather than the metal poor result found by
Jones et al. (2002).

\begin{figure}
\begin{center}
\includegraphics [width=88mm, height=50mm, angle=00]{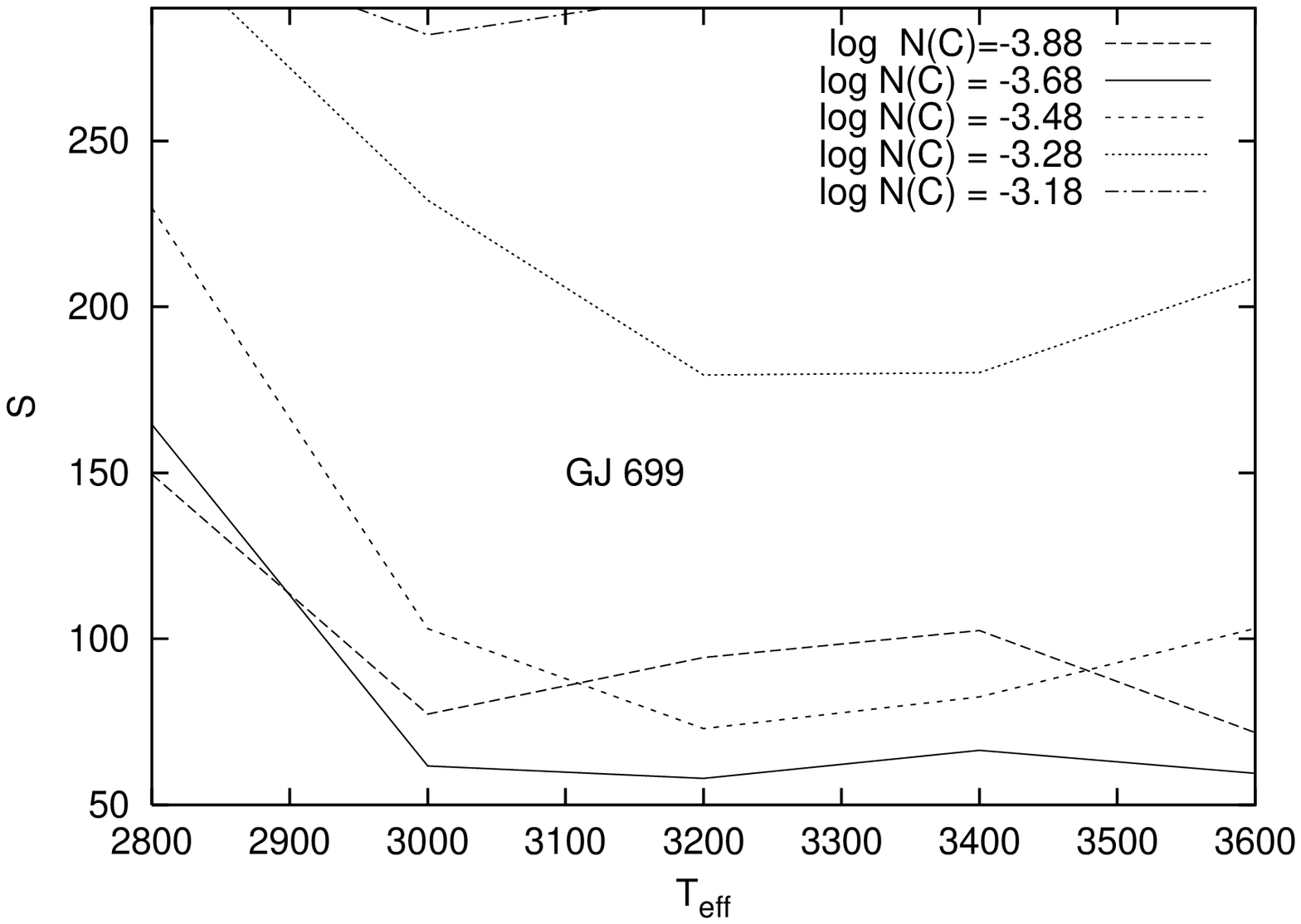}
\includegraphics [width=88mm, height=50mm, angle=00]{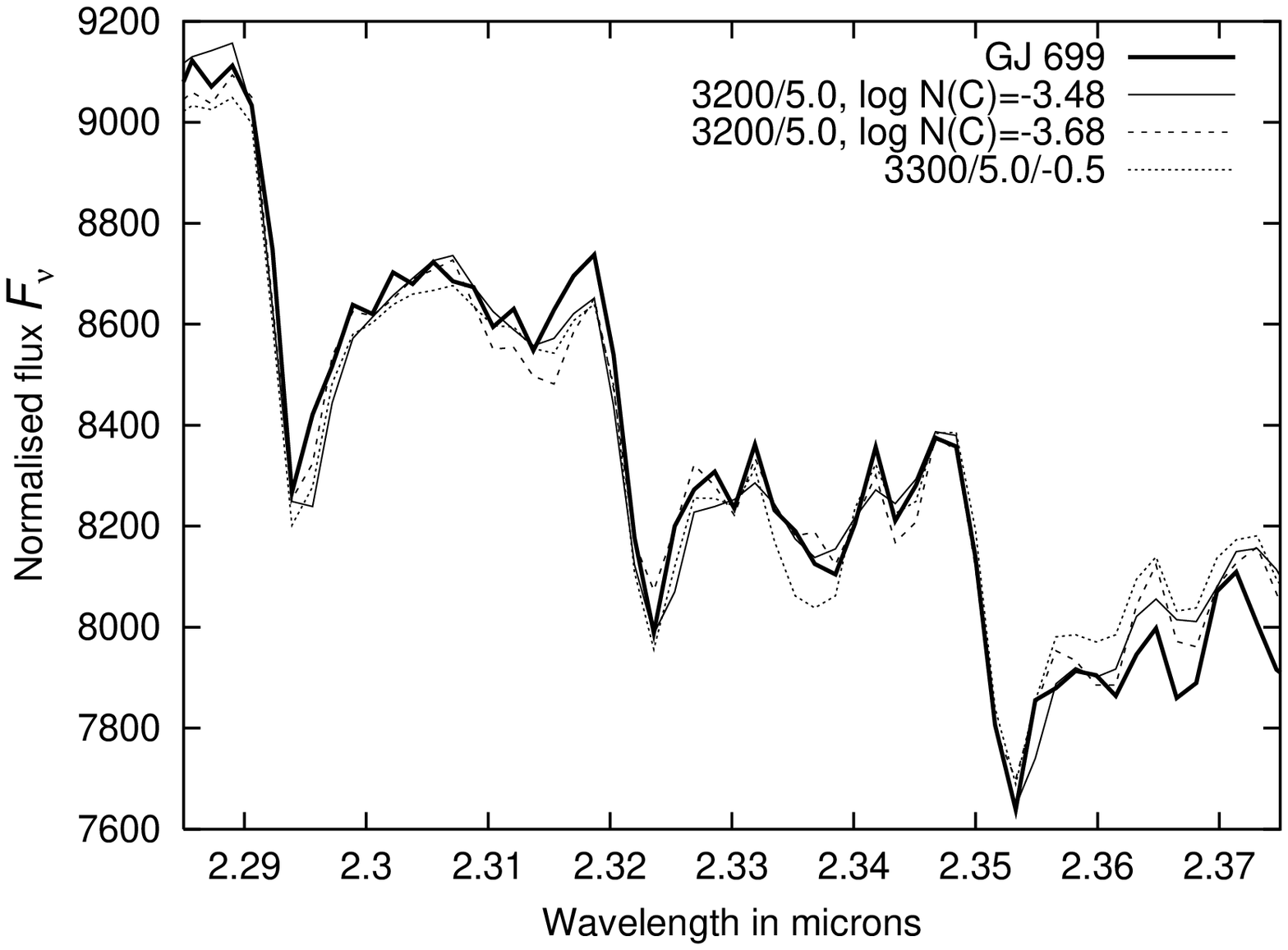}
\end{center}
\caption[]{\label{_GJ699_} The same as in Fig. \ref{_VB10_}, but
for GJ699. Synthetic spectra computed for 3400 and 3600 K
synthetic spectra are shown.}
\end{figure}

{\large\bf GJ699}.
The observed spectrum of GJ699 (M4V) is of lower resolution and
taken from the setup used in Jones et al. (1994). The
details in observed spectrum are best fit by metal-poor (log N(C)
= -3.68; i.e., [C] = -0.2) synthetic spectra of \Tef = 3200 K (see fig.
\ref{_GJ699_}). The same best fit temperature is also found for
the solar abundances log N(C)= -3.48, ([C] =0) case.
However, a carbon-deficient case log N(C) = -3.68 ([C]=-0.2)
provides lower $S$. We note that Jones et al. (2002) found an
effective temperature \Tef = 3300 K and metallicity [M/H] = --
0.5 ([O] = -0.5).
On the other hand, our fits with
log N(C) = -3.88 ([C] = -0.4) provide less
satisfactory fits. i.e fits of
higher $S$. We have probably reached the limit at which a single
metallicity parameter is appropriate for fits to the spectra
of M-dwarfs.

\begin{figure}
\begin{center}
\includegraphics [width=88mm, height=50mm, angle=00]{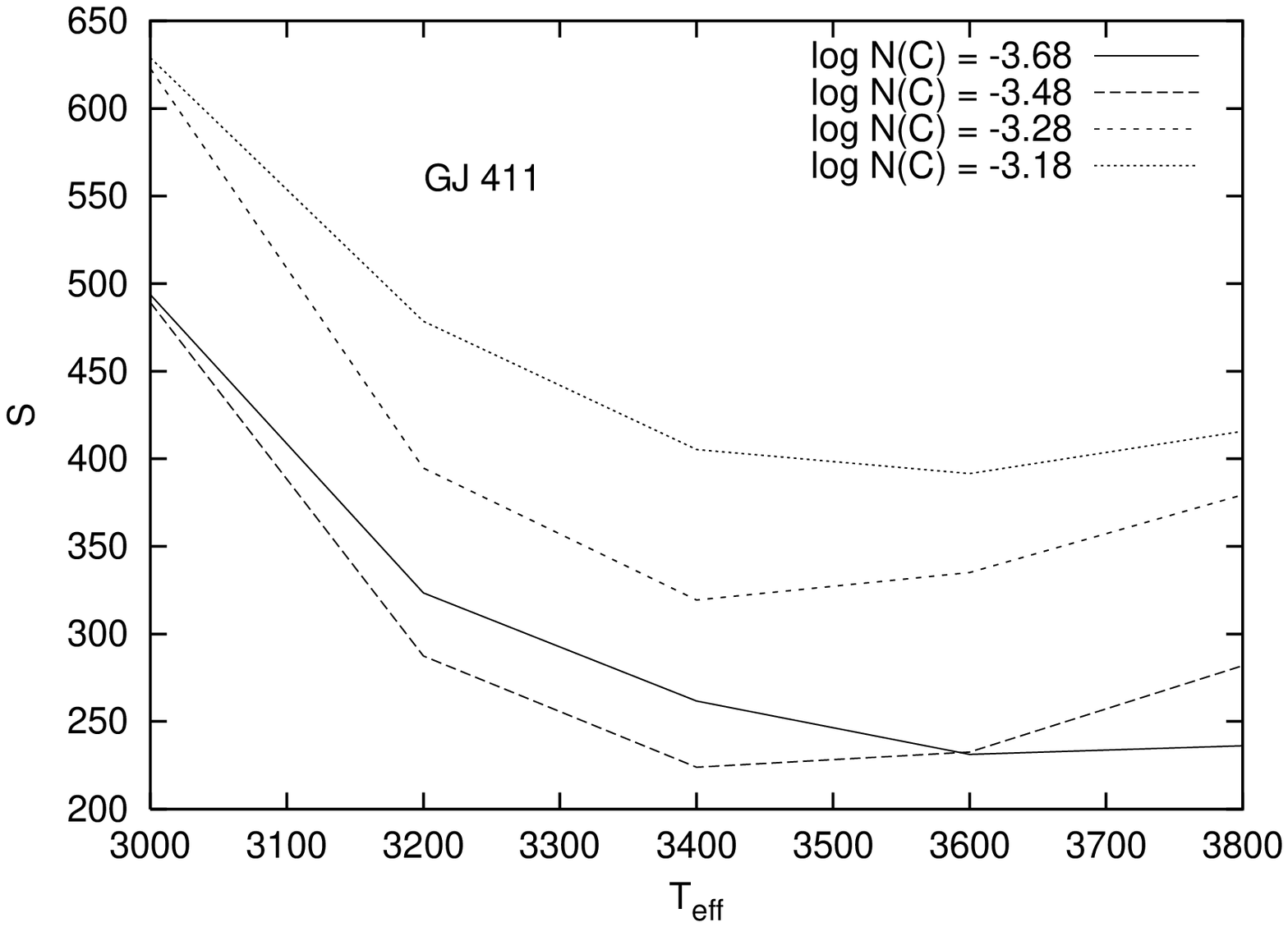}
\includegraphics [width=88mm, height=50mm, angle=00]{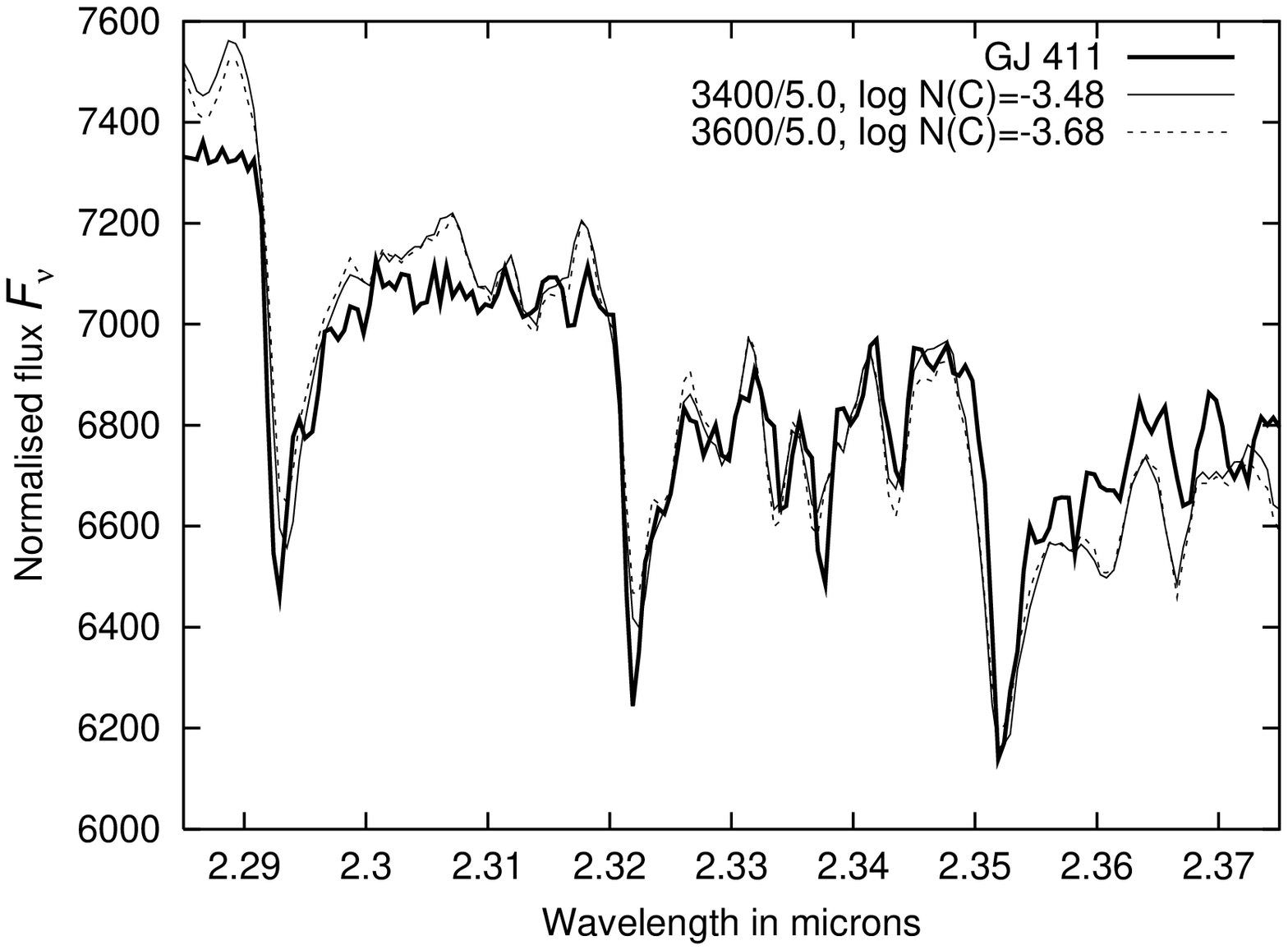}
\end{center}
\caption[]{\label{_GJ411_} The same as in Fig. \ref{_VB10_}, but
for GJ411 including synthetic spectra for 3400 and 3600 K.}
\end{figure}

{\large\bf GJ411} is the hottest dwarf in our sample. The fit to
the observed spectrum is shown in  Fig. \ref{_GJ411_}. Within
our observed spectral region the  intensities of observed  and computed
band heads of are fitted for \Tef = 3400 K and log N(C) = -3.48 
(Fig. \ref{_GJ411_}).
Although the relatively poorer description of
the M dwarf spectrum slope testifies to the presence of certain
problems in the definition of its physical characteristics:
abundances, homogeneity of atmospheres, chromospheres or other
reasons (see Section 5).

\begin {table*}
\begin {center}
\caption {\Tef of our sample for solar metallicity
compared with other methods: water
vapour (Jones et al. 2002), infrared flux method (Tsuji et al. 1997),
constant opacity (Jones et al. 1994),
model fit based on temperatures derived by Leggett et al. (2000, 2001).
The temperatures in italics are those derived for stars of similar 
spectral type.}
\begin {tabular} {ccccccccc}

\hline
\hline
\noalign{\smallskip}

 Object & Spectral & \Tef   & \Tef& \Tef   & \Tef& \Tef   \\
        & Type & This work & Water vapour & IRFM & Constant opacity
        & Model fit \\
\hline
\noalign{\smallskip}

 GJ~411 & M2V & 3400 & - & 3510 & 3471 & {\it 3450} \\
 GJ~699 & M4V &3200 & 3300 & 3210 & 3095 & 3100 \\
 GJ~406 & M6V & 2800 & 3000 & 2800 & 2670 & 2600 \\
 VB10 & M8 & 2600 &- & 2250 & 2506 & {\it 2250} \\
 LHS2924 & M9 & 2600 &- & 2120 & 2219 & {\it 2125} \\

\hline
\end {tabular}
\end {center}
\end {table*}

\subsubsection{Isotopic ratio \CDC in M dwarf atmospheres}

The temperatures in M dwarf interiors are relatively low (T $<$ 10$^6$ K),
so that hydrogen burning is only possible in pp chain
reactions and the hydrogen burning lifetimes are rather long
(e.g. Laughlin et al. 1997).
Thus despite the large convective envelopes extending down to 
the hydrogen burning region in a short timescales,
carbon isotope abundances cannot be changed even in
Hubble-scale times. M dwarfs are expected to
preserve their initial carbon isotopes abundance from their
time of formation.
The situation is very different for massive (M $>$
0.8 M$_{\odot}$) stars, which burn hydrogen in CNO reactions
and are observed to change their isotopic abundances of C,N,O.
The determination of the \CDC ratio in the
atmospheres of M dwarfs might 
provide important constraints on the theory of the evolution of the
Galaxy.



\begin{figure}
\begin{center}
\includegraphics [width=88mm, height=50mm, angle=00]{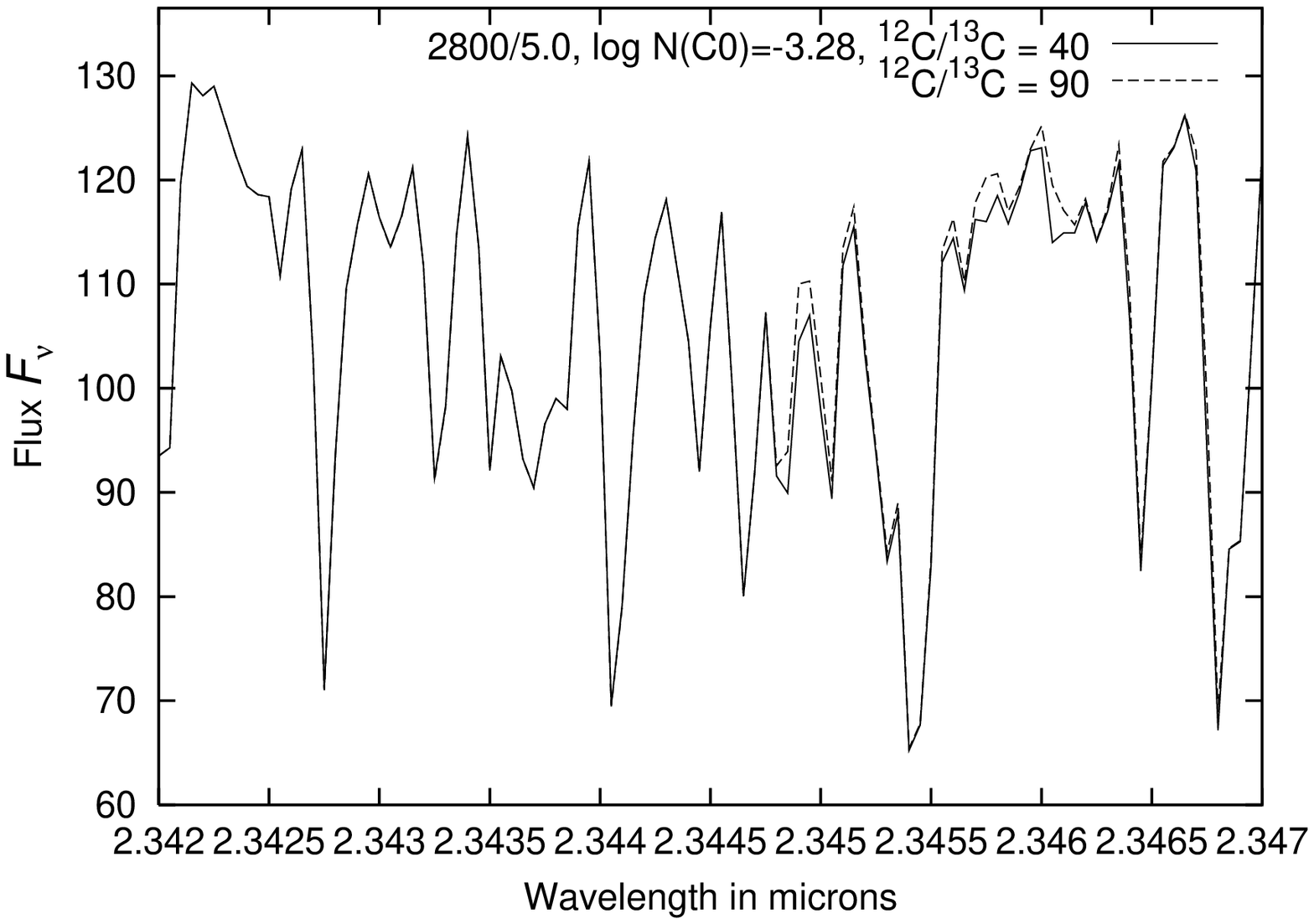}
\includegraphics [width=88mm, height=50mm, angle=00]{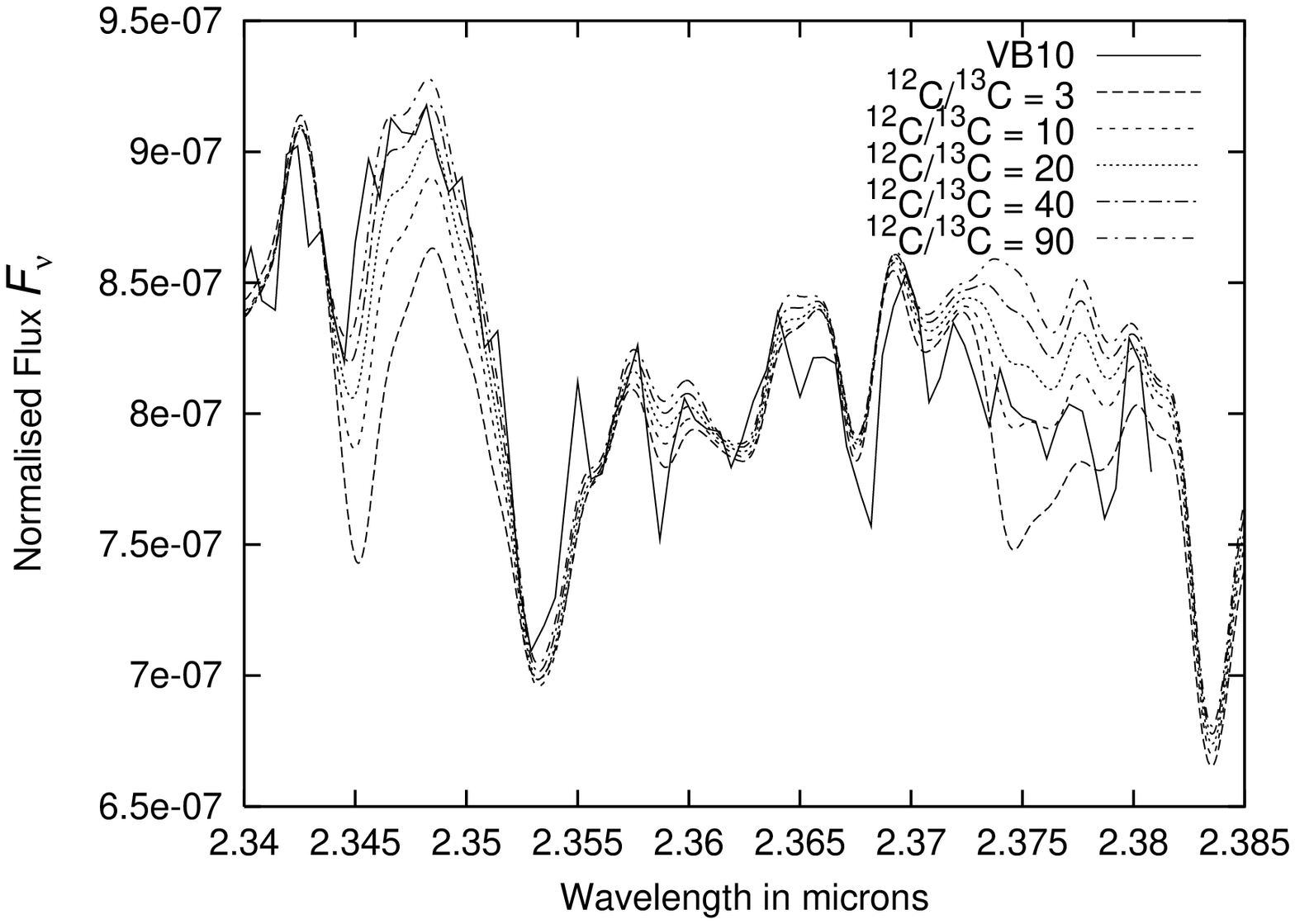}
\end{center}
\caption[]{\label{_CC1_}
Top: Spectra computed for model atmosphere 2800/5.0, log N(C)=-3.28,
for \CDC = 3 to 90 at a resolution of around 46000.
Bottom:
Fits of computed spectra with different \CDC to VB10 observed
spectrum.}
\end{figure}

We investigated the determination of the carbon isotopic ratio
in the atmospheres of our observed M dwarfs.
The analysis of the weak \CCC  bands in our comparatively
low-resolution spectra of dwarfs is rather difficult (e.g., Fig.
\ref{_CC1_} for VB10). 
The CO band head at 2.345
is mimiced by \HHO absorption. Moreover, the overall shape of the
energy distribution at these wavelengths is governed by \HHO
absorption, Fig. \ref{_H2O_CO_}. From the fit of our spectra
for VB10 we can estimate the lower limit \CDC $>$ 15. For more
accurate determination using this feature it will be necessary
to use spectra of better
resolution ($>$ 50000 for \CDC in range 40 --- 90, e.g., upper 
panel of Fig. \ref{_CC1_}).
The other band of \CCC at 2.375 $\mu$m provides a more sensitive
tool for \CDC determination in M-dwarf spectra.
From the fit of the observed band to our models we find \CDC $\sim$ 10.  
Unfortunately our confidence in this fit is weakened because our data
do not extend beyond 2.38 $\mu$m. To be sure we are measuring 
this feature we need to have data beyond 2.38 $\mu$m. 
For the other M dwarfs the 
fits are not so good so it is necessary to confirm this result with 
spectra of higher resolution and signal-to-noise.
Nevertheless,
from the fit of our spectra we can estimate the lower limit \CDC $>$ 10 
or all stars.


\subsubsection{Bands of the first overtone of CO from 4.3 to 4.6 microns}

\begin{figure}
\begin{center}
\includegraphics [width=88mm, height=50mm, angle=00]{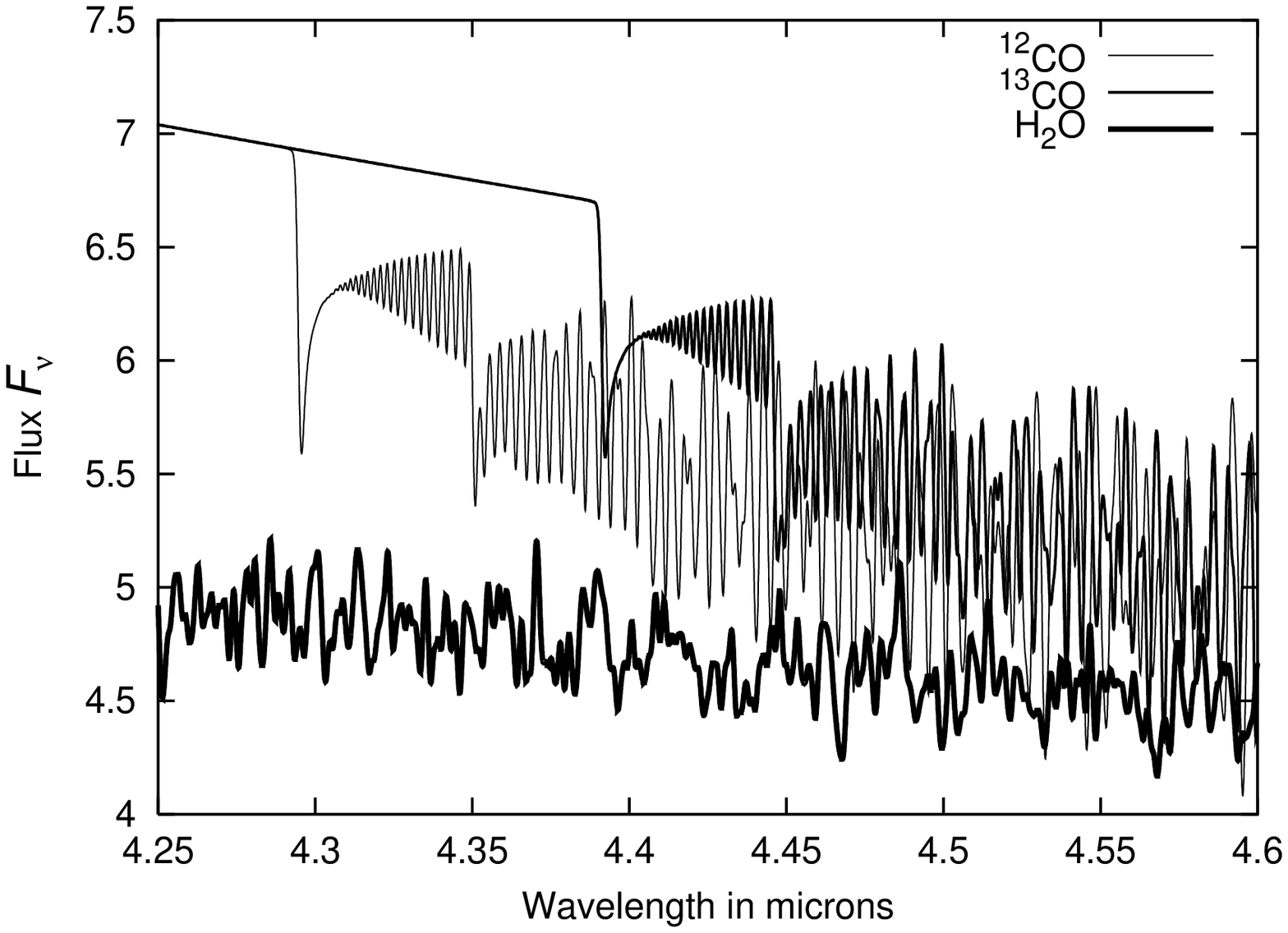}
\includegraphics [width=88mm, height=50mm, angle=00]{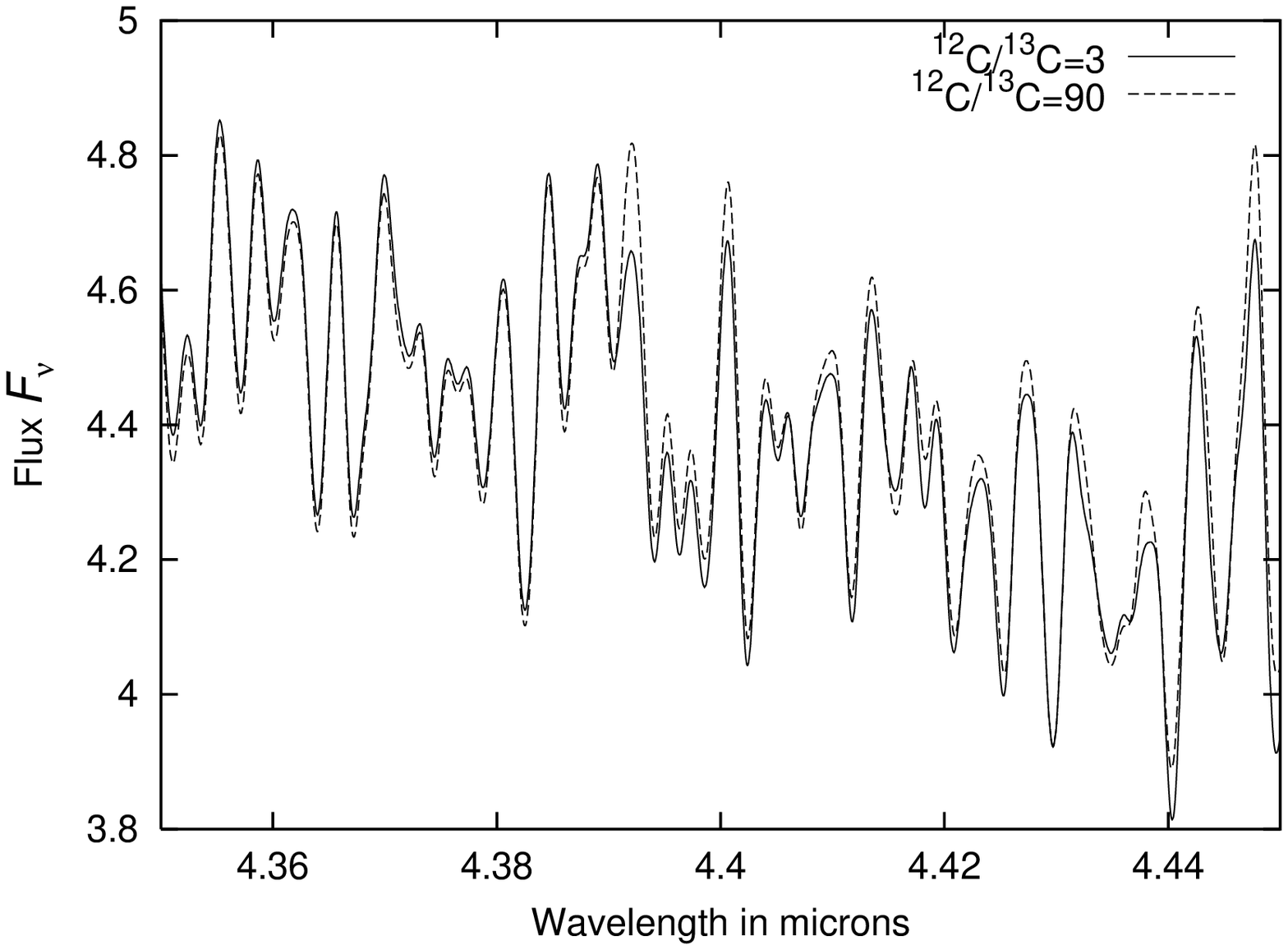}
\end{center}
\caption[]{\label{_45_} Top: relative strength of \CC , \HHO and
\CCC bands in theoretical spectrum of M-dwarf 2800/5.0/0. and a
combined spectra computed for the case of \CC + \HHO and \CC +
\CCC + \HHO absorption. Bottom: synthetical spectra computed for
2800/5.0
   model atmospheres and different \CDC. The resolution $R$ = 4600 was
   adopted.}
\end{figure}

As with the $\Delta\nu$ = 2 CO bands at 2.3$\mu$m it might
be feasible to pursue the
\CDC ratio using the $\Delta\nu$=1 CO bands at 4.4$\mu$m.
A theoretical picture of formation of the absorption features
in the region of the bands of the first overtone of CO is shown
in Fig. \ref{_45_}.
Again, as for the $^{13}$CO feature at 2.345 $\mu$m, 
lines of \HHO dominate in region (see also Fig.
\ref{_H2O_CO_}). Bands of $^{12}$CO (and $^{13}$CO) are
actually of lower intensities relative to water.
The band heads of $^{13}$C fall such that their heads
lie in the tails of $^{12}$C bands, which are
stronger. Therefore the determination of  
the \CDC ratio for $\Delta\nu$ = 1 CO bands is also problematic.


\section {Discussion}

Our effective temperature determinations are based primarily on \CC lines
but also affected by \HHO lines. In
Table 2 it can be seen that we find similar
temperatures to those found for M dwarfs for \HHO lines alone.  
These results support 
the expected quality of the \CC and  \HHO line lists.

In general, we find \Tef of our dwarfs with internal errors of around
$\pm$ 150 K. Table 2 shows \Tef derived for M dwarfs by different
methods. It illustrates indicates that whilst there is some agreement
for the earlier type M dwarfs, the temperatures we derive for
late spectral types are much too hot. 
This discrepancy arises because in the
atmospheres of red dwarfs of spectral class M6 and later (Tsuji,
Ohnaka \& Aoki 1996; Jones \& Tsuji 1997) appreciable amounts of
dust forms (e.g., Tsuji 2002)
that can absorb and/or scatter radiation.
This dust changes the structure of model atmospheres of
M dwarfs, and, hence, their spectra. The outermost layers of late
M dwarf atmospheres increase due to appearance there of the
``dusty'' opacity. In general, we get the much more sophisticated
picture of formation of late dwarf spectra in the ``dusty''
atmospheres in comparison with ``non-dusty'' case.
For the case of our fits
the dust in cooler stars causes the CO bands to be relatively
weaker and thus best-fit by a high temperature, non-dusty model.
Analysis of our data to determine reliable temperatures for
VB~10 and LHS~2924
thus requires the use of dusty models. Such models are now in
preparation (Pavlenko et al., in preparation).

NLTE (Non Local Thermal Equilibrium) effects provide another possible source of error. Carbon et al. (1976) investigated NLTE
effects in CO lines of the fundamental systems formed in the
atmospheres of stars of giant late-type spectral classes. However,
they used a rather simple model of rotational-vibrational levels
of CO. Furthermore they considered the model atmospheres of red
giants, where the densities are much lower, than in M dwarf
atmospheres. Nevertheless, they showed the possibility of NLTE
effects in CO lines, formed on depths $ \tau_{1{\mu}m} < 10^{-4}$.
CO lines in our cases are strong, they are formed deep in the
atmosphere (due to the low opacities in continuum). The formation
of absorption lines of CO occurs in M dwarfs atmospheres on a
background of strong  \HHO absorption. The additional absorption
should strengthen the thermalisation of the radiation field on CO
lines. The temperature structure of M dwarf atmospheres is
smoother, in comparison to the giants. Indeed, their structure is
governed by convection, which ``smoothes'' the temperature
contrast between photosphere and upper layers, which are more
pronounced in atmospheres of red giants. Atmospheres of cool
dwarfs are denser, than giants which strengthens the
thermalisation of the radiation field. All these factors should
reduce NLTE effects in CO lines formed in M dwarf atmospheres.
More precise estimations are possible and would be desirable when
making comparisons with higher resolution data though we consider
to be beyond the scope of the current work.

Our computations were carried out for model atmospheres of solar
abundance. Such abundances should be prevalent for the majority M
dwarfs in the vicinity of solar system. 
Although, our sample of M dwarfs
is rather small, we have found some deviation in carbon abundances
from the solar case. A more refined solution should be made using 
self-consistent abundances of oxygen and nitrogen.

The model atmospheres used in this work  were computed in
the framework of the classical assumptions by Hauschildt et al.
(1999). Energy transfer in atmospheres is provided by radiation
field and convection. The reliability of the computed model atmosphere
is determined by both the completeness of the opacity
sources as well as the model of convective processes(Ludwig et al. 2002). 
Absorption by \HHO
bands determines the opacity in the infrared spectra of red stars.
However, the existing line lists are not sufficiently complete
(e.g., Jones et al. 2002).
The CO line list used in this work was computed for solar work and
so is believed to be more than adequate for the lower energy
states accessed in M dwarf atmospheres (Goorvitch 1994). Overall
we do not consider that incompleteness in the \HHO line lists 
will cause temperatures derived from CO bands to change by more than 150 K.

The spectra presented here can not be used to determine reliable
\CDC ratios. Nonetheless we propose that with a slightly different
wavelength coverage than observed for this study, the $^{13}$CO
feature at 2.375 $\mu$m should be a strong constraint on the
\CDC ratio. For the other $^{13}$CO feature we investigated
the situation is complicated by \CCC bands forming
amongst strong \HHO absorption. One way of dealing with such
'contamination' would be to obtain higher resolution data.
In the upper panel of Fig. 9 we show synthetic spectra
at a resolution of around 46000, here even at low \CDC ratios 
the $^{12}$CO and $^{13}$CO bands can be distinguished.
We have also investigated the possible use
of the 4.4-4.6 micron region.  However, this region does not appear
to improve the possibility of determining \CDC.

\section{Acknowledgments}
This work is based on observations obtained at the United Kingdom Infrared
Telescope (UKIRT), which is operated by the Joint Astronomy Centre on
behalf of the UK Particle Physics and Astronomy Research
Council.
We thank David Schwenke and David Goorvitch for
providing AMES H$_2$O
and CO databases in digitial form.
We are also grateful to the anonymous referee for a number of useful
suggestions which significantly improved this manuscript.
This work was partially supported by a PPARC visitors grant to
the Astrophysics Research Institute, Liverpool John Moores University.
YPs studies are partially
supported by a Small Research Grant from American Astronomical
Society. 

\end{document}